\documentclass{vgtc}                          

\graphicspath{{figures/}{pictures/}{images/}{./}} 

\usepackage{times}                     

\usepackage{tabu}                      
\usepackage{booktabs}                  
\usepackage{lipsum}                    
\usepackage{mwe}                       

\usepackage{mathptmx}                  

\onlineid{0}

\vgtccategory{Research}

\vgtcinsertpkg

\title{Decoupling Data and Tooling in Interactive Visualization}

\author{Jan Simson\thanks{e-mail: jan.simson@lmu.de}\\ %
     \parbox{1.4in}{\scriptsize \centering LMU Munich \\ Munich Center for Machine Learning (MCML)}}

\teaser{
  \centering
  \includegraphics[width=\linewidth]{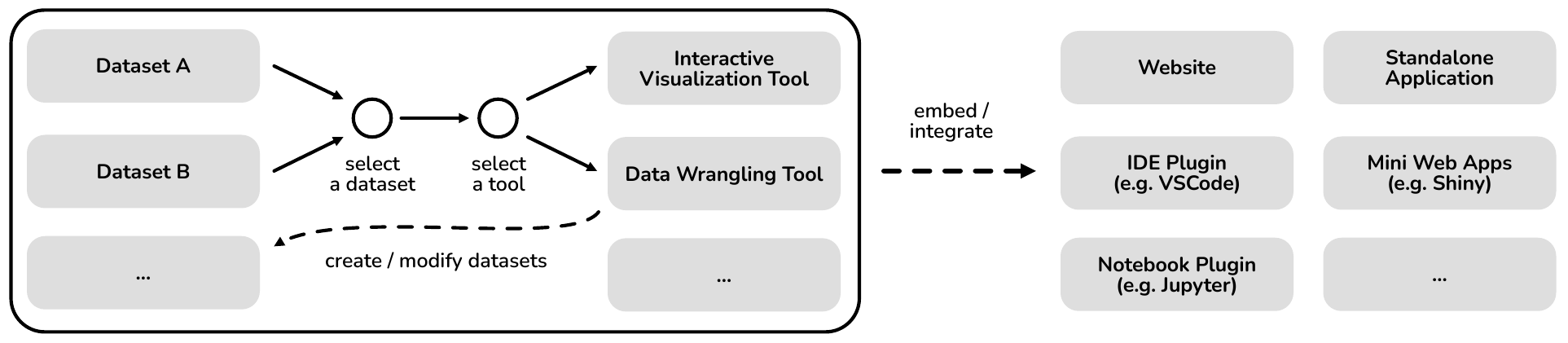}
  \caption{Illustration of the proposed abstraction. Separating data handling and loading from interactive visualization tools allows for easy combination of tools and embedding in new environments. Any particular tool only needs to be aware of incoming data, its own state and optionally outgoing data, simplifying the development of new tools while making them more powerful.}
  \label{fig:framework}
}

\abstract{Interactive data visualization is a major part of modern exploratory data analysis, with web-based technologies enabling a rich ecosystem of both specialized and general tools. However, current visualization tools often lack support for transformation or wrangling of data and are forced to re-implement their own solutions to load and ingest data. This redundancy creates substantial development overhead for tool creators, steeper learning curves for users who must master different data handling interfaces across tools and a degraded user experience as data handling is usually seen as an after-thought.

We propose a modular approach that separates data wrangling and loading capabilities from visualization components. This architecture allows visualization tools to concentrate on their core strengths while providing the opportunity to develop a unified, powerful interface for data handling. An additional benefit of this approach is that it allows for multiple tools to exist and be used side by side. We demonstrate the feasibility of this approach by building an early prototype using web technologies to encapsulate visualization tools and manage data flow between them.

We discuss future research directions, including downstream integrations with other tooling, such as IDEs, literate programming notebooks and applications, as well as incorporation of new technologies for efficient data transformations. We seek input from the community to better understand the requirements towards this approach.
} 

\keywords{Interactive visualization, visualization tooling, data manipulation.}

\begin{document}


\firstsection{Introduction}

\maketitle

Data analysis rarely follows a linear path. This becomes evident when examining established frameworks such as the Cross-Industry Standard Process for Data Mining (CRISP-DM) \cite{chapman2000crispdm} or the data science life cycle described in the PCS-framework / veridical data science \cite{yu2020veridical}, both of which highlight the cyclic nature of data science processes. Both frameworks recognize that analysts cycle continuously between data preparation, manipulation, and exploratory analysis phases. Interactive data visualization anchors this exploratory process, enabling pattern recognition, hypothesis formation, and effective communication of findings \cite{heer2012interactive}.

Web technologies have become the dominant platform for developing and deploying interactive visualization tools. The widespread availability of web browsers, combined with powerful libraries, has democratized visualization creation and fostered a diverse ecosystem spanning general-purpose chart builders to highly specialized tools for specific use-cases.

Despite these advances, a critical usability challenge persists. Most visualization tools require input data in specific formats—whether tidy data structures or particular schemas, necessitating complex transformations even for basic exploratory plots. Current web-based tools address this requirement by developing custom data loading and parsing workflows. While data loading solutions can be polished, as demonstrated by tools like RAWGraphs \cite{maurei2017rawgraphs}, they are created in isolation and not every developer has the resources available to prioritize user experience during data loading. Beyond just the loading of data, few tools allow for the manipulation, reshaping or transformation of data, rather expecting data to be supplied in the expected format.

This isolated development approach creates two significant problems. First, it forces developers to repeatedly solve the same data handling challenges, diverting resources from core visualization innovation. Second, it fragments the user experience across tools, requiring analysts to learn new data import and manipulation interfaces for each visualization they want to create. Critical data wrangling tasks—cleaning, transforming, and reshaping data—often receive inadequate support or require switching between different applications, creating inefficient workflows that interrupt analytical thinking. Even within the visualization process it is common to switch between tools \cite{bigelow2017iterating}.

We argue for an architectural shift in web-based visualization tool design. By abstracting data loading and transformation logic into a separate, reusable layer, visualization tools can focus exclusively on their visualization strengths. This decoupling would streamline new tool development while enabling a standardized, powerful data wrangling interface that serves an entire ecosystem of visualizations. Such a framework would facilitate seamless integration across diverse environments, including browser-based editors (e.g. Visual Studio Code or Rstudio), standalone desktop applications built with web technologies (e.g. Tauri or Electron), literate programming notebooks (e.g. Jupyter or Quarto), websites and small interactive web-apps (e.g. Shiny or Streamlit).

In this work, we describe a new approach to decouple handling of data from visualization tools, implement a proof-of-concept prototype and ask for input from the community to understand the demands for a comprehensive and long-term solution.

\section{Prototype}

We develop a proof-of-concept prototype to explore the feasibility of the discussed approach. This prototype supports multiple views, grouped into data manipulation and visualization. For data manipulation we implement a simple data loader, a data viewer, an interactive spreadsheet to modify data by hand, a data wrangler\footnote{\url{https://observablehq.com/@observablehq/data-wrangler}.} and a pivot table. For data visualization, we implemented support for Sanddance \cite{drucker2015unifying}, Voyager \cite{wongsuphasawat2017voyager}, Perspective\footnote{\url{https://github.com/finos/perspective}}, kepler.gl\footnote{\url{https://kepler.gl/}} and an interactive plotly.js editor\footnote{\url{https://github.com/plotly/react-plotly.js}}. The prototype is available online at \url{https://data-studio.simson.io/}.

The architecture's core principle involves encapsulating individual visualization tools within \textit{iframes}, managed by a host application that serves as the central hub for data management and tool coordination. Source code is available at \url{https://github.com/jansim/data-studio/}.

\subsection{Challenges}

A key challenge in the creation of this prototype is the large heterogeneity of dependencies and build pipelines between different visualization tools. As the web ecosystem evolved over time, so have the different build systems in it and both build pipelines as well as dependencies are often incompatible between different tools. We address this issue, by building each tool separately and running them in isolated contexts using \textit{iframes}. Data is passed between the main frame and the tools on demand.

Second, as the web is not originally designed for working with tabular data it lacks a standardized format and different visualization tools expect data in different formats, usually a combination of JavaScript objects and arrays. This challenge is addressed, by allowing tools to request data from the main frame in different formats and applying tool-specific transformations for uncommon formats if necessary.

\begin{figure}[tb]
 \centering
 \includegraphics[width=\columnwidth]{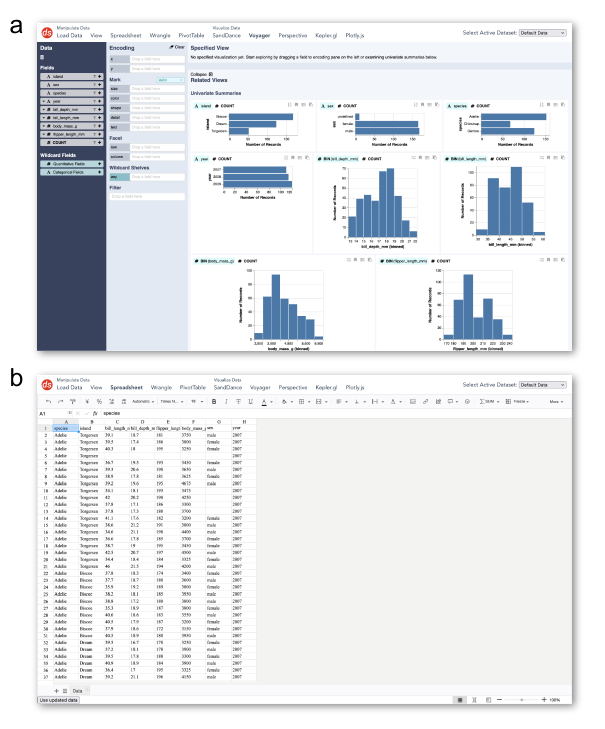}
 \caption{A screenshot of the prototype implementation, illustrating the \textit{Voyager} \cite{wongsuphasawat2017voyager} view (a) for exploratory data analysis and the \textit{Spreadsheet} view (b) to examine and modify data.}
 \label{fig:screenshots}
\end{figure}

\section{Discussion}

This work and particular the prototype successfully demonstrate the viability of separating data handling from visualization in web-based interactive visualization tools. However, it is important to note that the prototype is indeed a prototype and is still lacking in features for usage in a production settings. Indeed the goal with this work is to better understand which pieces are missing to enable the development of a more comprehensive solution.

Emerging technologies like the WebAssembly (WASM) port of DuckDB \cite{raasveldt2019duckdb} could provide a more efficient way of handling data in a future version of the system, potentially in conjunction with large language models (LLMs) to generate SQL code and allow for data wrangling using natural language. The system could also serve as a standard component for interactive notebooks, IDE plugins, and desktop analytical applications.

Success in this endeavor requires community input to establish standards, address technical challenges, and ensure the approach meets real-world needs. Through collaborative development, we can create a solution that has the potential to enhance the ecosystem of web-based data analysis tools.


\bibliographystyle{abbrv-doi}

\bibliography{template}
\end{document}